\documentclass[apjl]{emulateapj}
\usepackage{apjfonts}
\usepackage{textcomp}
\usepackage{amsfonts}
\usepackage{amssymb}

\newcommand{\td}{\textdegree}
\newcommand{\be}{\begin{enumerate}}
\newcommand{\ee}{\end{enumerate}}

\shorttitle{Observations of IC 443 with VERITAS}
\shortauthors{Acciari et al.}

\begin{document}


\title{Observation of Extended VHE Emission from the Supernova Remnant IC 443 with VERITAS}


\author{
V.~A.~Acciari\altaffilmark{1},
E.~Aliu\altaffilmark{2},
T.~Arlen\altaffilmark{3},
T.~Aune\altaffilmark{4},
M.~Bautista\altaffilmark{5},
M.~Beilicke\altaffilmark{6},
W.~Benbow\altaffilmark{1},
S.~M.~Bradbury\altaffilmark{7},
J.~H.~Buckley\altaffilmark{6},
V.~Bugaev\altaffilmark{6},
Y.~Butt\altaffilmark{8},
K.~Byrum\altaffilmark{9},
A.~Cannon\altaffilmark{10},
O.~Celik\altaffilmark{3,\dag},
A.~Cesarini\altaffilmark{11},
Y.~C.~Chow\altaffilmark{3},
L.~Ciupik\altaffilmark{12},
P.~Cogan\altaffilmark{5},
P.~Colin\altaffilmark{13},
W.~Cui\altaffilmark{14},
M.~K.~Daniel\altaffilmark{7,\ddag},
R.~Dickherber\altaffilmark{6},
C.~Duke\altaffilmark{15},
V.~V.~Dwarkadas\altaffilmark{16},
T.~Ergin\altaffilmark{8},
S.~J.~Fegan\altaffilmark{3,\S},
J.~P.~Finley\altaffilmark{14},
G.~Finnegan\altaffilmark{13},
P.~Fortin\altaffilmark{17,\S},
L.~Fortson\altaffilmark{12},
A.~Furniss\altaffilmark{4},
D.~Gall\altaffilmark{14},
K.~Gibbs\altaffilmark{1},
G.~H.~Gillanders\altaffilmark{11},
S.~Godambe\altaffilmark{13},
J.~Grube\altaffilmark{10},
R.~Guenette\altaffilmark{5},
G.~Gyuk\altaffilmark{12},
D.~Hanna\altaffilmark{5},
E.~Hays\altaffilmark{18},
J.~Holder\altaffilmark{2},
D.~Horan\altaffilmark{19},
C.~M.~Hui\altaffilmark{13},
T.~B.~Humensky\altaffilmark{20,*},
A.~Imran\altaffilmark{21},
P.~Kaaret\altaffilmark{22},
N.~Karlsson\altaffilmark{12},
M.~Kertzman\altaffilmark{23},
D.~Kieda\altaffilmark{13},
J.~Kildea\altaffilmark{1},
A.~Konopelko\altaffilmark{24},
H.~Krawczynski\altaffilmark{6},
F.~Krennrich\altaffilmark{21},
M.~J.~Lang\altaffilmark{11},
S.~LeBohec\altaffilmark{13},
G.~Maier\altaffilmark{5},
A.~McCann\altaffilmark{5},
M.~McCutcheon\altaffilmark{5},
J.~Millis\altaffilmark{25},
P.~Moriarty\altaffilmark{26},
R.~A.~Ong\altaffilmark{3},
A.~N.~Otte\altaffilmark{4},
D.~Pandel\altaffilmark{22},
J.~S.~Perkins\altaffilmark{1},
M.~Pohl\altaffilmark{21},
J.~Quinn\altaffilmark{10},
K.~Ragan\altaffilmark{5},
L.~C.~Reyes\altaffilmark{27},
P.~T.~Reynolds\altaffilmark{28},
E.~Roache\altaffilmark{1},
H.~J.~Rose\altaffilmark{7},
M.~Schroedter\altaffilmark{21},
G.~H.~Sembroski\altaffilmark{14},
A.~W.~Smith\altaffilmark{9},
D.~Steele\altaffilmark{12},
S.~P.~Swordy\altaffilmark{20},
M.~Theiling\altaffilmark{1},
J.~A.~Toner\altaffilmark{11},
L.~Valcarcel\altaffilmark{5},
A.~Varlotta\altaffilmark{14},
V.~V.~Vassiliev\altaffilmark{3},
S.~Vincent\altaffilmark{13},
R.~G.~Wagner\altaffilmark{9},
S.~P.~Wakely\altaffilmark{20},
J.~E.~Ward\altaffilmark{10},
T.~C.~Weekes\altaffilmark{1},
A.~Weinstein\altaffilmark{3},
T.~Weisgarber\altaffilmark{20},
D.~A.~Williams\altaffilmark{4},
S.~Wissel\altaffilmark{20},
M.~Wood\altaffilmark{3},
B.~Zitzer\altaffilmark{14}
}

\altaffiltext{1}{Fred Lawrence Whipple Observatory, Harvard-Smithsonian Center for Astrophysics, Amado, AZ 85645, USA}
\altaffiltext{2}{Department of Physics and Astronomy and the Bartol Research Institute, University of Delaware, Newark, DE 19716, USA}
\altaffiltext{3}{Department of Physics and Astronomy, University of California, Los Angeles, CA 90095, USA}
\altaffiltext{4}{Santa Cruz Institute for Particle Physics and Department of Physics, University of California, Santa Cruz, CA 95064, USA}
\altaffiltext{5}{Physics Department, McGill University, Montreal, QC H3A 2T8, Canada}
\altaffiltext{6}{Department of Physics, Washington University, St. Louis, MO 63130, USA}
\altaffiltext{7}{School of Physics and Astronomy, University of Leeds, Leeds, LS2 9JT, UK}
\altaffiltext{8}{Harvard-Smithsonian Center for Astrophysics, 60 Garden Street, Cambridge, MA 02138, USA}
\altaffiltext{9}{Argonne National Laboratory, 9700 S. Cass Avenue, Argonne, IL 60439, USA}
\altaffiltext{10}{School of Physics, University College Dublin, Belfield, Dublin 4, Ireland}
\altaffiltext{11}{School of Physics, National University of Ireland, Galway, Ireland}
\altaffiltext{12}{Astronomy Department, Adler Planetarium and Astronomy Museum, Chicago, IL 60605, USA}
\altaffiltext{13}{Department of Physics and Astronomy, University of Utah, Salt Lake City, UT 84112, USA}
\altaffiltext{14}{Department of Physics, Purdue University, West Lafayette, IN 47907, USA }
\altaffiltext{15}{Department of Physics, Grinnell College, Grinnell, IA 50112-1690, USA}
\altaffiltext{16}{Department of Astronomy and Astrophysics, University of Chicago, Chicago, IL, 60637}
\altaffiltext{17}{Department of Physics and Astronomy, Barnard College, Columbia University, NY 10027, USA}
\altaffiltext{18}{N.A.S.A./Goddard Space-Flight Center, Code 661, Greenbelt, MD 20771, USA}
\altaffiltext{19}{Laboratoire Leprince-Ringuet, Ecole Polytechnique, CNRS/IN2P3, F-91128 Palaiseau, France}
\altaffiltext{20}{Enrico Fermi Institute, University of Chicago, Chicago, IL 60637, USA}
\altaffiltext{21}{Department of Physics and Astronomy, Iowa State University, Ames, IA 50011, USA}
\altaffiltext{22}{Department of Physics and Astronomy, University of Iowa, Van Allen Hall, Iowa City, IA 52242, USA}
\altaffiltext{23}{Department of Physics and Astronomy, DePauw University, Greencastle, IN 46135-0037, USA}
\altaffiltext{24}{Department of Physics, Pittsburg State University, 1701 South Broadway, Pittsburg, KS 66762, USA}
\altaffiltext{25}{Department of Physics, Anderson University, 1100 East 5th Street, Anderson, IN 46012}
\altaffiltext{26}{Department of Life and Physical Sciences, Galway-Mayo Institute of Technology, Dublin Road, Galway, Ireland}
\altaffiltext{27}{Kavli Institute for Cosmological Physics, University of Chicago, Chicago, IL 60637, USA}
\altaffiltext{28}{Department of Applied Physics and Instrumentation, Cork Institute of Technology, Bishopstown, Cork, Ireland}
\altaffiltext{*}{Corresponding author: humensky@uchicago.edu}
\altaffiltext{\dag}{Currently at N.A.S.A./Goddard Space-Flight Center, Code 661, Greenbelt, MD 20771, USA}
\altaffiltext{\ddag}{Currently at Department of Physics, Durham University, Durham, DH1 3LE, UK}
\altaffiltext{\S}{Currently at Laboratoire Leprince-Ringuet, Ecole Polytechnique, CNRS/IN2P3, F-91128 Palaiseau, France}



\begin{abstract}
We present evidence that the very-high-energy (VHE, $E > 100\ \textrm{GeV}$) gamma-ray emission coincident with the supernova remnant IC 443 is extended.  IC 443 contains one of the best-studied sites of supernova remnant/molecular cloud interaction and the pulsar wind nebula CXOU J061705.3+222127, both of which are important targets for VHE observations.  VERITAS observed IC 443 for 37.9 hours during 2007 and detected emission above $300\ \textrm{GeV}$ with an excess of 247 events, resulting in a significance of 8.3 standard deviations ($\sigma$) before trials and 7.5 $\sigma$ after trials in a point-source search.  The emission is centered at $6^{\textrm{h}}16^{\textrm{m}}51^{\textrm{s}}+22\textrm{\td}30'11''$ (J2000) $\pm 0.03\textrm{\td}_{stat} \pm 0.08\textrm{\td}_{sys}$, with an intrinsic extension of $0.16\textrm{\td} \pm 0.03\textrm{\td}_{stat} \pm 0.04\textrm{\td}_{sys}$.  The VHE spectrum is well fit by a power law ($dN/dE = N_0 \times (E/\textrm{TeV})^{-\Gamma}$) with a photon index of $2.99 \pm 0.38_{stat} \pm 0.3_{sys}$ and an integral flux above 300 GeV of $(4.63 \pm 0.90_{stat} \pm 0.93_{sys}) \times 10^{-12}\ \textrm{cm}^{-2}\ \textrm{s}^{-1}$.  These results are discussed in the context of existing models for gamma-ray production in IC 443.
\end{abstract}





\section{Introduction}
IC 443 (G189.1 +3.0) is one of the most thoroughly studied supernova remnants (SNRs) and remains one of the clearest examples of an SNR interacting with molecular clouds. IC 443 has a double-shell structure in the optical and radio \citep{Braun1986}.  The northeast shell's interaction with the H\footnotesize\ II\normalsize\ region S249 places IC 443 at a distance of $1.5\ \textrm{kpc}$ \citep{Fesen1984a}. A second shell is expanding into a less-dense region to the southwest.  The age of IC 443 remains uncertain, with various estimates placing it in the range $\sim$$3\textrm{-}30\ \textrm{kyr}$ \citep{Petre1988a,Troja2008a,Lee2008a,Chevalier1999a}.

According to \citet{Cornett1977} and \citet{Dickman1992a}, a molecular cloud encircles the remnant.  OH maser emission from the central and southeast regions indicates interaction with the foreground portion of the cloud \citep{Claussen1997a,Hewitt2006a}.  The total mass of the cloud is $\sim$$10^4\ \textrm{M}_{\odot}$ \citep{Torres2003a}. \citet{Dickman1992a} and \citet{Lee2008a} estimate that $\sim$$500\textrm{-}2000\ \textrm{M}_{\odot}$ of cloud material have been directly perturbed by the SNR shock.  

IC 443's X-ray emission is primarily thermal and peaked towards the interior of the northeast shell \citep{Petre1988a,Troja2006a}, placing IC 443 in the mixed-morphology class of SNRs.  Observations by Chandra \citep{Olbert2001a,Gaensler2006a} and XMM \citep{Bocchino2001a} of the southern edge of the shell have resolved a pulsar wind nebula (PWN), CXOU J061705.3+222127, that may be the compact remnant of IC 443's progenitor. Although the direction of motion of the neutron star, inferred from the morphology of the X-ray PWN, does not point back to the center of the shell, asymmetric expansion of the shell and distortion of the PWN tail by turbulence in the ambient medium could be exaggerating the apparent misalignment \citep{Gaensler2006a}. No pulsations have been detected at any wavelength.

The centroid of the unidentified EGRET source 3EG J0617+2238 is located near the center of the IC 443 shell, but the limited angular resolution of EGRET makes it impossible to resolve the $\sim$$45'$ extension of the remnant complex.  The EGRET source has a spectral index of $-2.01 \pm 0.06$ over the band $100\ \textrm{MeV - }30\ \textrm{GeV}$ and an integral flux above $100\ \textrm{MeV}$ of $5 \times 10^{-7}\ \textrm{cm}^{-2}\ \textrm{s}^{-1}$ \citep{Hartman1999a}. AGILE (AGL J0617+2236) and the Fermi Gamma-ray Space Telescope (0FGL J0617.4+2234) have recently reported gamma-ray detections with position and flux consistent with EGRET \citep{Pittori2009a,Abdo2009a}.  IC 443 was detected in very-high-energy (VHE, $E > 100\ \textrm{GeV}$) gamma rays by MAGIC \citep{Albert2007a} and VERITAS \citep{Humensky2007a}. MAGIC J0616+225 is a point-like source coincident with the densest part of the molecular cloud and the OH maser emission observed therein.  It has a power-law spectrum described by $(1.0 \pm 0.2_{stat} \pm 0.35_{sys}) \times 10^{-11} (E/0.4\ \textrm{TeV})^{-3.1 \pm 0.3_{stat} \pm 0.2_{sys}}\ \textrm{TeV}^{-1}\ \textrm{cm}^{-2}\ \textrm{s}^{-1}$ in the range $0.1\textrm{-}1.6\ \textrm{TeV}$.  In this paper, deep VHE gamma-ray observations of IC 443 with VERITAS are reported that reveal extended emission coincident with the molecular cloud structure.  

\section{Observations and Analysis}


VERITAS \citep{Holder2006a} consists of four $12\textrm{-m}$ imaging atmospheric Cherenkov telescopes located at an altitude of $1268\ \textrm{m}$ a.s.l. at the Fred Lawrence Whipple Observatory in southern Arizona, USA (31\td\ 40' 30'' N, 110\td\ 57' 07'' W). Each telescope is equipped with a 499-pixel camera of $3.5\textrm{\td}$ field of view.  The array, completed in the spring of 2007, is sensitive to a point source of 1\% of the steady Crab Nebula flux above $300\ \textrm{GeV}$ at $5\ \sigma$ in less than 50 hours at 20\td\ zenith angle.

VERITAS observed IC 443 during two epochs in 2007.  Data set 1 was acquired with three telescopes during the commissioning phase in February and March of 2007. The PWN location of $6^{\textrm{h}}17^{\textrm{m}}5.3^{\textrm{s}}+22\textrm{\td}21'27''$ (J2000) was tracked in wobble mode, in which the source is displaced by $0.5\textrm{\td}$ from the center of the field of view \citep{Fomin1994a}. Approximately equal amounts of data were acquired in each of four offset directions. Additional wobble observations with four telescopes were taken during October and November of 2007, using the centroid location determined from the earlier data, $6^{\textrm{h}}16^{\textrm{m}}52.8^{\textrm{s}}+22\textrm{\td}33'0''$ (J2000).  Application of standard quality-selection cuts on weather conditions and hardware and rate stability resulted in 16.8 hours live time in data set 1 and 20.9 hours in data set 2, with mean zenith angles of 20\td\ and 17\td, respectively.  Data set 2 is used for the spectral analysis.  In this data set, 0.8 hours were taken with a non-standard array trigger configuration, and these runs are excluded from the spectral analysis.

The data are analyzed following standard procedures described in \citet{Cogan2007a} and \citet{Daniel2007a}.  Showers are reconstructed for events in which at least two telescope images passed pre-selection criteria: number of pixels in the camera image $>$ 4, number of photoelectrons in the image $> 75$, and distance of the image centroid from the camera center $<$ 1.43\td.  These criteria impose an energy threshold\footnote{Peak of the differential counting rate for a Crab Nebula-like spectrum.} of $300\ \textrm{GeV}$.

The event-selection cuts are optimized for weak ($\sim$$1\textrm{-}5\%$ of the Crab Nebula flux) point sources using data taken on the Crab Nebula. Cosmic-ray rejection is performed using the \textit{mean-scaled width} (selecting events within the range 0.05-1.24) and \textit{mean-scaled length} (selecting events within the range 0.05-1.40) parameters, as described in \citet{Konopelko1995a}. The ring-background model \citep{Aharonian2005a} is used to study the morphology and the reflected-region model \citep{Aharonian2001a} is used to determine the spectrum.  Regions of radius 0.4\td\ around the source location and 0.3\td\ around two bright stars are excluded from the background estimation.  All results presented here have been verified by an independent calibration and analysis chain; in particular, the morphology was cross-checked using a template background model \citep{Rowell2003a} because it is sensitive to a different combination of systematic effects than the ring-background model; this is especially relevant for fields that contain bright stars.

One systematic issue particular to this data set is the presence of two bright stars in the IC 443 field: Eta Gem (V band magnitude 3.31), located 0.53\td\ from the PWN; and Mu Gem (V band magnitude 2.87), located 1.37\td\ from the PWN. The high flux of optical photons from the stars requires that several photomultiplier tubes (PMTs) in each camera be turned off during the observations and produces higher noise levels in the signals of nearby PMTs.  These effects reduce the exposure in the vicinity of the stars and degrade the angular resolution.  A high telescope-multiplicity requirement, described below, was chosen for studying morphology in order to make the direction reconstruction more robust against these effects.  In order to verify that the presence of a bright star in the field does not produce a false excess, a $4.1\textrm{-hr}$ exposure was taken on a sky field that contained a magnitude-2.7 star (36 Eps Boo) but lacked any likely gamma-ray sources.  No excess is observed at the location equivalent to IC 443 in the 36 Eps Boo field.

\section{Results}
Here an updated, standard point-source analysis of the combined data set (1 \& 2) is presented, in which a maximum significance of 8.3 standard deviations ($\sigma$) before trials (7.5 $\sigma$ post trials, accounting for a blind search over the region enclosed by the shell of IC 443) is found at $06^{\textrm{h}}16^{\textrm{m}}49^{\textrm{s}}\textrm{+22\td} 28'30''\ (\textrm{J2000})$ and a significance of 6.8 $\sigma$ at the location of MAGIC J0616+225.  The significance is calculated according to equation 17 of \citet{Li1983a}.  Table~\ref{tbl-1} summarizes the results of this analysis at the location of maximum significance, listing the counts falling within the point-source integration radius of 0.112\td\ ($on$), the counts integrated in a background ring spanning radii 0.6-0.8\td\ ($off$), the ratio of $on$ exposure to $off$ exposure ($\alpha$), and the resulting number of excess counts and significance.

\subsection{Morphology}
\begin{figure}[h]
\begin{center}
\includegraphics[width=0.86\textwidth]{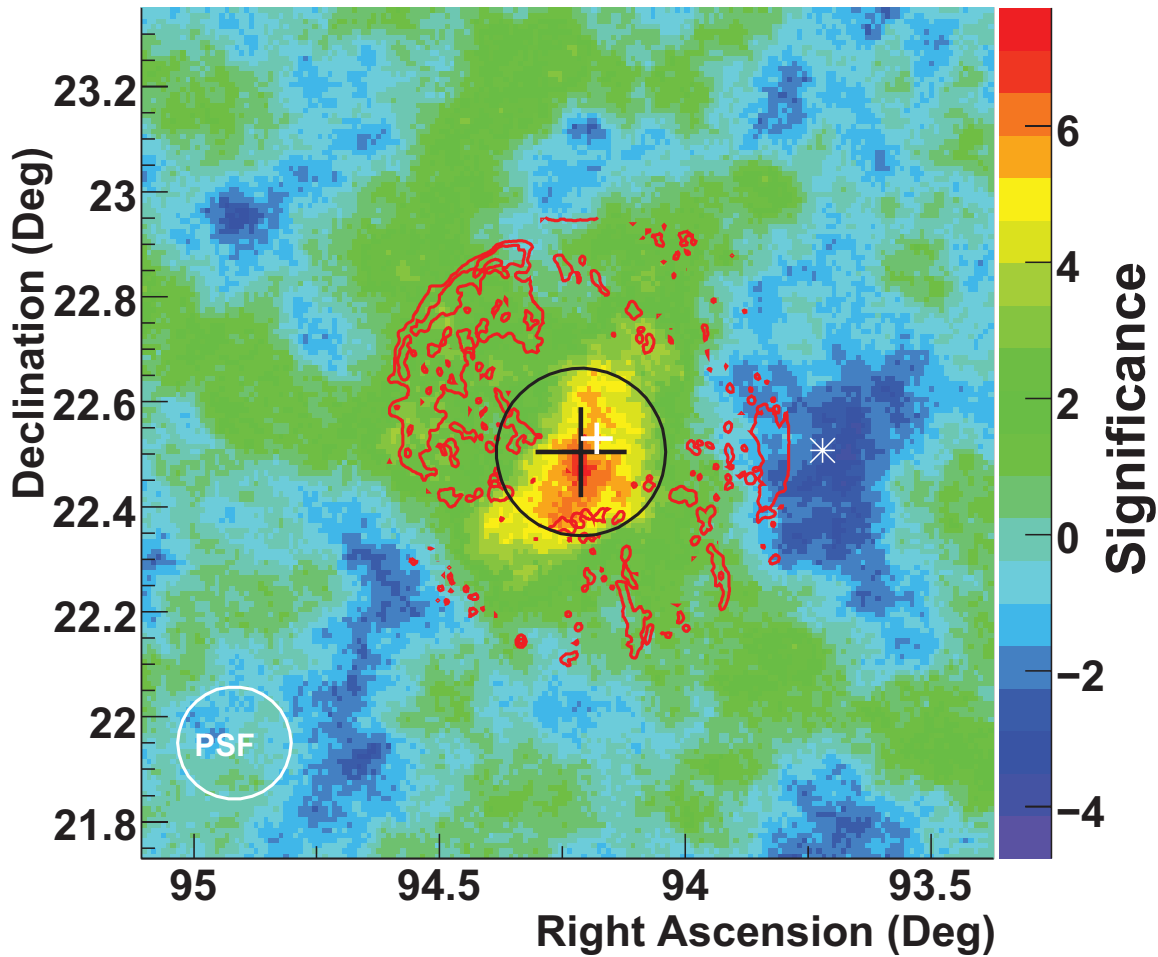}
\end{center}
\caption{Significance map for the IC 443 field.  The  black circle indicates the one-sigma angular extension (see text). The black cross-hair indicates the centroid position and its uncertainty (statistical and systematic added in quadrature), and the white cross-hair likewise indicates the position and uncertainty of MAGIC J0616+225 \citep{Albert2007a}.  Red contours: optical intensity \citep{McLean2000a}.  White star: location of the star Eta Gem.  The white circle indicates the PSF of the VERITAS array.}\label{fig:sigmapRBM}
\end{figure}

Figure~\ref{fig:sigmapRBM} shows the significance map for the IC 443 field.  To study the source morphology, an integration radius of 0.112\td\ is used and the best-reconstructed gamma rays are selected by requiring that an event must have all 3 images and all 4 images surviving the pre-selection cuts in data sets 1 and 2, respectively.  This requirement increases the analysis energy threshold by $\sim$15\% and reduces the excess by $\sim$40\%.  The centroid and intrinsic extension of the excess are characterized by fitting an azimuthally symmetric two-dimensional Gaussian, convolved with the PSF\footnote{The PSF is characterized as a sum of two, two-dimensional Gaussians describing a narrow core and a broader tail.  It is determined from data taken on the Crab Nebula, which is a point source at these energies \citep{Albert2008a}.} of the instrument, to an acceptance-corrected uncorrelated excess map with a bin size of 0.05\td.  The PSF has a 68\% containment radius of 0.11\td.  The centroid is located at $06^{\textrm{h}}16^{\textrm{m}}51^{\textrm{s}}\textrm{+22\td} 30'11''\ (\textrm{J2000})\ \pm 0.03\textrm{\td}_{stat} \pm 0.08\textrm{\td}_{sys}$, consistent with the MAGIC position.  The extension derived in this fashion is $0.16\textrm{\td} \pm 0.03\textrm{\td}_{stat} \pm 0.04\textrm{\td}_{sys}$.  The difference in extension between this work and the point-like source detected by \citet{Albert2007a} can be explained by the difference in angular resolution and sensitivity between VERITAS and MAGIC, the latter of which has an angular resolution of $\sim$$0.15\textrm{\td}$ for 68\% containment and a sensitivity above $200\ \textrm{GeV}$ of 2\% of the Crab Nebula flux in 50 hours \citep{Albert2008a}.

\subsection{Spectrum}
\begin{figure}[h]
\begin{center}
\includegraphics [width=0.86\textwidth]{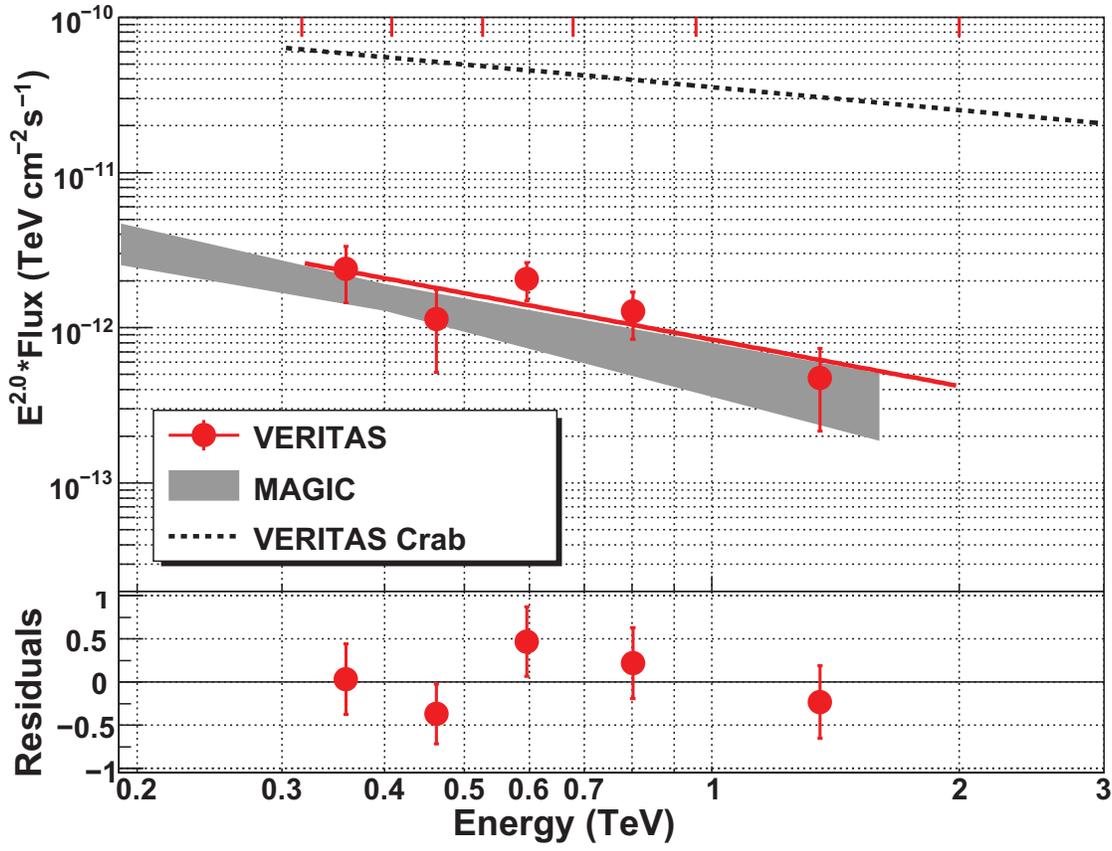}
\end{center}
\caption{Spectrum of IC 443, scaled by $E^2$.  Red points are the VERITAS spectrum (red tick marks along top indicate bin edges).  The red line is a power-law fit (see text), with residuals to the fit plotted in the lower box.  The MAGIC spectrum is indicated by the gray band and extends down to $90\ \textrm{GeV}$.  VERITAS error bars and MAGIC error band reflect statistical errors only.  The Crab spectrum (V. Acciari et al. 2009, in preparation) is shown as a dashed line for comparison.}\label{fig:spectrum}
\end{figure}

The threshold for the spectral analysis is $300\ \textrm{GeV}$; the energy resolution is less than $\sim$20\% at $1\ \textrm{TeV}$.  The photon spectrum, integrated within a radius of 0.235\td, is shown in Figure~\ref{fig:spectrum}.  The photon spectrum is well fit ($\chi^2/\textrm{ndf} = 3.1/3$) by a power law $dN/dE = N_0 \times (E/\textrm{TeV})^{-\Gamma}$ in the range $0.317\textrm{-}2.0\ \textrm{TeV}$, with a normalization of $(8.38 \pm 2.10_{stat} \pm 2.50_{sys}) \times 10^{-13}\ \textrm{TeV}^{-1}\ \textrm{cm}^{-2\ }\textrm{s}^{-1}$ and an index of $2.99 \pm 0.38_{stat} \pm 0.30_{sys}$.  The integral flux above $300\ \textrm{GeV}$ is $(4.63 \pm 0.90_{stat} \pm 0.93_{sys}) \times 10^{-12}\ \textrm{cm}^{-2}\ \textrm{s}^{-1} $ (3.2\% of the Crab Nebula flux), consistent within errors with the spectrum reported by MAGIC \citep{Albert2007a}.

\section{Discussion and Conclusions}
\begin{figure}[h]
\begin{center}
\includegraphics[width=0.86\textwidth]{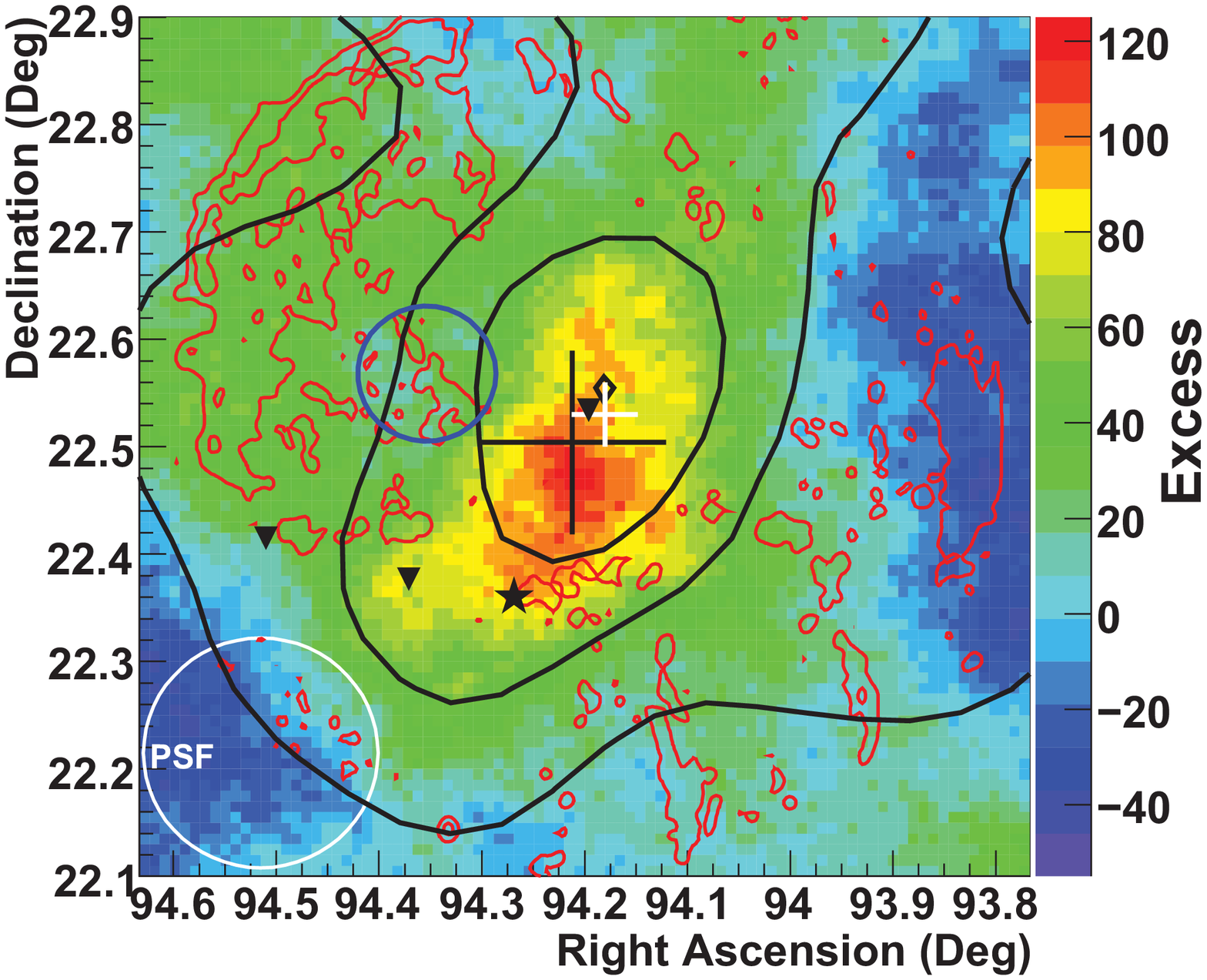}
\end{center}
\caption{Inner $0.8\textrm{\td}$ of the acceptance-corrected excess map for the IC 443 field. Markers and contours follow Figure~\ref{fig:sigmapRBM}, with several additions.  Thick black contours: CO survey \citep{Huang1986a}; black star: PWN CXOU J061705.3+222127 \citep{Olbert2001a}; open blue circle: 95\% confidence radius of 0FGL J0617.4+2234 \citep{Abdo2009a}; and filled black triangles: locations of OH maser emission (Claussen et al. 1997; J. W. Hewitt, private communication).}\label{fig:mwlmap}
\end{figure}

Figure~\ref{fig:mwlmap} shows the inner $0.8\textrm{\td}$ of the excess map and places the VHE emission in a multi-wavelength context. The TeV centroid is $\sim 0.15\textrm{\td}$ from the position of the PWN and $\sim 0.03\textrm{\td}$ from the nearby, bright maser emission coincident with clump G of \citet{Huang1986a}.  Two additional weaker regions of maser emission lie along the southern rim of the remnant, east of the PWN \citep{Hewitt2006a}.  The VHE emission overlaps the foreground molecular cloud.  While the Fermi source 0FGL J0617.4+2234 is displaced from the centroid of the VHE emission by $\sim$0.15\td, this is consistent at the 95\% level with the combined errors quoted by Fermi and VERITAS.

The VHE gamma-ray emission observed by VERITAS and MAGIC is offset from the location of the PWN by $\sim$$10\textrm{-}20$ arcmin, commensurate with other offset PWNe such as HESS J1825-137 \citep{Aharonian2006a}. The emission is consistent with a scenario in which the VHE emission arises from inverse Compton scattering off electrons accelerated early in the PWN's life.  Such a scenario was presented by \citet{Bartko2008a}.  Under the assumption that the VHE emission is associated with IC 443 at a distance of $1.5\ \textrm{kpc}$, the luminosity in the energy band $0.3\textrm{-}2.0\ \textrm{TeV}$ is $4 \times 10^{32}\ \textrm{erg s}^{-1}$.  The spin-down luminosity of the pulsar has been estimated as $\sim$$10^{36}\ \textrm{erg s}^{-1}$ \citep{Olbert2001a,Bocchino2001a} and $5 \times 10^{37}\ \textrm{erg s}^{-1}$ \citep{Gaensler2006a}.  If the PWN association is correct, the VHE luminosity in the $0.3\textrm{-}2.0\ \textrm{TeV}$ band is less than $\sim$$0.04\textrm{\%}$ of the spin-down luminosity, well within the $\sim$$0.01\textrm{-}10\%$ range of observed VHE efficiencies for other PWNe \citep{Gallant2007a}.  
Likewise, synchrotron cooling is unlikely to rule out an offset PWN scenario.  The lifetime $\tau$ of a synchrotron-emitting electron with energy $E$ is $\tau(E) \sim 1.3 \times 10^7 (B/\mu\textrm{G})^{-2}(E/1\ \textrm{GeV})^{-1}\ \textrm{kyr}$ \citep{Gaisser1998a}.  Assuming the TeV photons are produced by $\sim$$20\textrm{-TeV}$ electrons in a magnetic field of $\sim$$5\ \mu\textrm{G}$ (comparable to typical interstellar fields), synchrotron cooling is only just beginning to become important even if the age of IC 443 is as high as $\sim$$30\ \textrm{kyr}$.  Note that in the nebula-powered scenario presented by \citet{Bartko2008a}, the GeV emission is assumed to be unresolved pulsar emission.  The displacement of 0FGL J0617.4+2234 from the PWN location argues against that scenario.

Alternatively, Figure~\ref{fig:mwlmap} can be interpreted within a scenario of hadronic cosmic-ray acceleration and subsequent interaction with the molecular cloud, which would provide a high density of target material for the production of VHE gamma rays.  The correlation of the VHE emission with the molecular cloud is natural within this scenario.  The low velocity of the SNR shock implied by the presence of maser emission \citep{Hewitt2006a} indicates that the shock is in its radiative phase in this region and cosmic-ray acceleration is most likely now inefficient.  Thus, the cosmic rays accelerated during the shock's earlier propagation have diffused into the remnant and the upstream region, including the molecular cloud.  The steep VHE spectrum can be explained either as a low maximum energy to which particles were accelerated prior to the shock hitting the cloud, or an energy-dependent rate of diffusion of cosmic rays out of the cloud \citep{Aharonian1996a}.  \citet{Zhang2008a} model the remnant as evolving partially within and partially outside the cloud, and find that a hadronic interpretation can explain the VHE emission.  However, \citet{Torres2008a} point out that in the \citet{Zhang2008a} model, the GeV and TeV emission ought to be spatially coincident; it is not yet clear whether they are.  \citet{Torres2008a} present an alternative picture in which cosmic rays escape from the SNR shock and diffuse toward the cloud, similar to what may be occurring in the Galactic Ridge \citep{Aharonian2006b} and the region surrounding the SNR W 28 \citep{Aharonian2008a}.  This picture can naturally accommodate spatially separated GeV and TeV emission by positing clouds at different distances from the SNR shock.  \citet{Marrero2009a} extend this work to make predictions for the Fermi Gamma-ray Space Telescope (Fermi) and find that Fermi should see a movement of the centroid  towards the TeV position as the photon energy increases, along with a gradual spectral softening.  The combination of GeV and TeV observations may also provide additional constraints on the diffusion coefficient and the distance between the SNR and the molecular cloud.  

Further GeV and TeV gamma-ray observations will help to clarify the nature of the particle acceleration associated with IC 443.  The $\sim$0.15\td\ separation between the centroids of 0FGL J0617.4+2234 and VER J0616.9+2230, though not statistically significant, may hint at an energy-dependent morphology.  Such a morphology will be explored with future observations by VERITAS, and has the potential to discriminate between the scenarios outlined above.  A break in the energy spectrum between the GeV and TeV bands is required whether or not 0FGL J0617.4+2234 and VER J0616.9+2230 are directly related.  Future observations by VERITAS will improve the statistics above $100\ \textrm{GeV}$, anchoring the measurement of the VHE cut-off to the gamma-ray emission.  The combined spectrum from $100\ \textrm{MeV}$ to several TeV will provide powerful constraints on the emission mechanism(s)~\citep{Gaisser1998a}.  

In summary, VERITAS observations of the composite supernova remnant IC 443 during 2007 have yielded a detection of an extended source with an index of $2.99 \pm 0.38_{stat} \pm 0.30_{sys}$ and a flux of $(4.63 \pm 0.90_{stat} \pm 0.93_{sys}) \times 10^{-12}\ \textrm{cm}^{-2}\ \textrm{s}^{-1} $ above $300\ \textrm{GeV}$. The location and flux are consistent with MAGIC J0616+225. This deep VERITAS observation reveals that the VHE emission is extended ($0.16\textrm{\td} \pm 0.03\textrm{\td}_{stat} \pm 0.04\textrm{\td}_{sys}$), with its brightest region coincident with the dense cloud material and maser emission.  The emission is offset from the nearby PWN CXOU J061705.3+222127, a viable source for the VHE emission.  If further VHE observations reveal an energy-dependent morphology, it may become possible to distinguish cleanly between scenarios related to the PWN and to hadronic cosmic rays interacting with the molecular cloud.



\acknowledgments

This research is supported by grants from the US Department of Energy, the US National Science Foundation, and the Smithsonian Institution, by NSERC in Canada, by Science Foundation Ireland, and by STFC in the UK. We acknowledge the excellent work of the technical support staff at the FLWO and the collaborating institutions in the construction and operation of the instrument. Some of the simulations used in this work have been performed on the Joint Fermilab - KICP Supercomputing Cluster.



{\it Facilities:} \facility{FLWO:VERITAS}.

\begin{table*}
\begin{center}
\caption{Analysis results for VERITAS observations of IC 443.\label{tbl-1}}
\begin{tabular}{lccccc}
\tableline\tableline
 & $on$ & $off$ & $\alpha$ & excess & significance \\
      &      &       &          &        & ($\sigma$)   \\
\tableline
data set 1 & 393 & 5152 & 0.0610 & 78.7 & 4.1  \\
data set 2 & 609 & 7331 & 0.0601 & 168.1 & 7.3  \\
combined   & 1002 & 12483 & 0.0604 & 247.5 & 8.3 \\
\tableline
\end{tabular}
\end{center}
\end{table*}


\end{document}